\documentclass{elsart}
\usepackage{epsfig}
\usepackage{rotate}
\usepackage{amssymb}
\usepackage{amsmath}

\newcommand\pair{e^+ e^-}

\newcommand\nuhd{\nu_h \rightarrow \gamma \nu}

\newcommand\nuh{\nu_h}
\newcommand\nus{\nu_h}
\newcommand\ee{e^+e^-}
\newcommand\nudecay{\nu_h \rightarrow \gamma \nu}
\newcommand\mix{|U_{\mu h}|^2}

\begin{document}
\begin{frontmatter}
\title{\boldmath New limits on radiative sterile neutrino decays from a search for  single photons in neutrino interactions.}
\begin{center}

S.N. Gninenko\\
\vskip0.5cm
{\em Instutute for Nuclear Research, Moscow}
\end{center}

\begin{abstract} 
It has been recently shown that excess events observed by  
the LSND and MiniBooNE neutrino experiments could be interpreted as a signal from  
the radiative decay of a heavy sterile neutrino $\nu_h$ produced in $\nu_\mu$ 
neutral current-like  neutrino interactions. 
If the $\nu_h$ exist, it would  be also produced by the $\nu_\mu$ beam from the CERN SPS  
in the neutrino beam line shielding. The  $\nu_h$'s  would penetrate the  shielding and be observed  through the $\nuhd$  decay  followed by the photon conversion into  $e^+e^-$ pair in the active target of the NOMAD detector. The $\nu_h$'s could be also produced in the 
 iron of the magnetic spectrometer of the CHORUS detector, located just in front of NOMAD. Considering these two 
 sources of $\nu_h$'s we  set new constraints on  $\nuh$ properties and exclude part of the 
 LSND/MiniBooNE  $\nu_h$ parameter space using  bounds on  single photons production in neutrino reactions recently  reported by the NOMAD collaboration. We find that broad bands in the parameter space are still open for  
 more sensitive searches for the $\nuh$ in future neutrino experiments.  
\end{abstract}
\begin{keyword} 
neutrino mixing, neutrino decay
\end{keyword}
\end{frontmatter}


Over the past 10 years there is a  puzzle of the 3.8 $\sigma$ event excess observed 
by the LSND collaboration \cite{lsndfin}. 
This excess originally interpreted as a signal from 
$\overline{\nu}_\mu \to \overline{\nu}_e$ 
 oscillations was not confirmed by further measurements from the similar KARMEN experiment
\cite{karmen}.  
The MiniBooNE experiment, designed to examine the LSND effect, 
 also  found no evidence  for $\nu_\mu \to \nu_e$ oscillations.
However,  an anomalous  excess of low energy electron-like events
in  quasi-elastic neutrino events 
 over the expected standard  
 neutrino interactions  has been observed \cite{minibnu2}.
 Recently, MiniBooNE has reported new  results from a search for
$\overline{\nu}_\mu \to \overline{\nu}_e$ oscillations \cite{minibnub}.
 An excess of events was  observed which has a small probability to be identified as  the 
background-only events. The data are found to be  consistent with $\overline{\nu}_\mu \to \overline{\nu}_e$ oscillations in the 0.1 eV$^2$
range and with the evidence for antineutrino oscillations from the LSND.

In the recent work \cite{sng} (see also \cite{sng1,gg,sngmu}) it has been shown  that these puzzling results could all be explained in a consistent way  by assuming  
the existence of a  heavy sterile neutrinos ($\nu_h$). The $\nu_h$ is created in  
$\nu_\mu$ {\em neutral-current} (NC) interactions and decay subsequently   into  
a photon and a lighter  neutrino $\nu$ in the  LSND and MiniBooNE detectors, 
but it cannot be produced in the KARMEN experiment  due to the high energy threshold.
The  $\nu_h$ could be Dirac or Majorana type.  The $\nu_h$ could decay  
{\em dominantly} into $\gamma \nu$ pair  if, for example,  there is a large enough  transition
magnetic moment between the $\nu_h$ and $\nu$ mass states.  
Assuming the $\nu_h$ is produced through mixing with $\nu_\mu$, 
the combined analysis of the LSND and MiniBooNe excess events suggests that 
 the $\nu_h$  mass, mixing strength, and lifetime are,  respectively,  in the range
 \begin{equation}
 40\lesssim m_h \lesssim 80~ \text{MeV},~ 10^{-3}\lesssim |U_{\mu h}|^2 \lesssim 10^{-2}, 10^{-11}\lesssim \tau_h\lesssim 10^{-9}~s.
 \label{param}
 \end{equation}
A detailed discussion of consistency of these values  with the constraints from
previous searches for heavy neutrinos \cite{pdg} is presented in \cite{sng}.
 Briefly, the mixing of \eqref{param} is not constrained by the limits from the 
  the most sensitive experiments searched for extra peaks in two-body $\pi, K$ decays \cite{pdg,kek}, because  the $\nu_h$ mass range of \eqref{param} is 
 outside of the kinematical limits for $\pi_{\mu 2}$ decays, and  not accessible to 
 $K_{\mu 2}$ experiments due to experimental resolutions.
The parameter space of \eqref{param} cannot be ruled out by the results of high energy  neutrino experiments, 
such as  NuTeV \cite{nutev} or CHARM \cite{charm1,charm2}, as they searched for $\nu_h$'s of higher   masses ($ m_h \gtrsim 200~ \text{MeV}$)  
 decaying into muonic  final states ($\mu \pi \nu,~\mu \mu \nu,~\mu e \nu, ..$) \cite{pdg}, which are not 
 allowed in our case. The best limits on $|U_{\mu h}|^2$  derived for the mass range \eqref{param} 
  from the search for $\nu_h \to e^+ e^- \nu$ decays in the PS191 experiment \cite{ps191}, 
 as well as the LEP bounds \cite{aleph},  are found to be compatible with \eqref{param} 
 assuming  the dominance of the $\nuh$ decay. New limits on  mixing $|U_{\mu h}|^2$  obtained by 
 using the recent results on precision measurements of the muon Michel parameters by the TWIST 
experiment \cite{twist} are also found to be consistent with \eqref{param}.
 Finally, the most stringent 
bounds on $\mix$ coming from the primordial nucleosynthesis and SN1987A considerations, 
as well as the limits from the atmospheric neutrino experiments,
are also evaded due to the short $\nu_h$ lifetime.

Recently, several constraints on the properties of the $\nu_h$ obtained from the muon capture on hydrogen \cite{mcp} and radiative decays of charged kaons \cite{koval,duk} have been reported. The new bounds, found to be in tension with \eqref{param}, are obtained under
assumption that the $\nu_h$, as a component of  the $\nu_\mu$, could be also produced in reactions induced by
$\nu_\mu$ charge-current (CC) interactions. However, one may argue that the heavy neutrino explanation of 
 the LSND/MiniBooNE anomaly  assumes that excess events are originated from the decays of $\nu_h$'s produced in a new    
$\nu_\mu$NC-like interactions, i.e. without charge muon emission \cite{sng}. 
Therefore, the constraints of Ref.'s \cite{mcp,koval,duk} can be evaded, if e.g. the new fermion $\nu_h$ has new dominant NC-like interactions, while its CC  reactions are suppressed.  

There are several possibilities for such scenario. For example, the $\nu_h$ can be produced 
by the Primakoff conversion, $\nu_\mu + Z \to \nu_h + Z$ of the muon neutrino in the electromagnetic field of nuclei due to 
dipole transition moment between the $\nu_h$ and $\nu_\mu$ states, with the 
subsequent decay $\nuhd$ due to the same interaction, see, e.g. Fig. 1 in  \cite{sgnk} and, 
also  \cite{proposal} for more detail discussions.
Another mechanisms, we are mostly interested in,  adopt new interaction to produce sterile neutrinos in 
muon neutrino scattering off nuclei. In the standard model, in our case of low energies, 
both $Z$- and $W$-boson contributions to neutrino scattering are naturally of the same order, because there is no hierarchy between $Z$- and $W$-boson masses. Would $W$-boson be much heavier, charge 
current contributions to neutrino scattering off nuclei are suppressed as compared 
to that of neutral current. Consequently, to diminish the meaning of charge 
current channel in searches for sterile neutrinos one can involve a similar hierarchy for the new interaction responsible for sterile neutrino production. Such a model based on  adding to the SM one new scalar doublet with the  same gauge quantum numbers as the SM Higgs field \cite{dsg} is discussed elsewhere 
\cite{proposal}. Similar scenario could arise also in the low-energy  superstring-inspired $E_6$ theories \cite{rizzo}.
In all these cases one does not need any sterile-active neutrino mixing at all to explain 
sterile neutrino appearance in muon neutrino beam traveling through a neutrino detector 
volume.   Therefore, it would be interesting to obtain direct experimental constraints on the such $\nu_h$ properties.  

 In this Letter we  study  new  constraints on  properties of $\nuh$'s from the measurements of the NOMAD experiment 
at the CERN WANF neutrino beam,  assuming the $\nu_h$'s  are   produced dominantly in new 
$\nu_\mu$NC-like  reactions.  
The present analysis as well as the experimental signature of the signal events are similar to those of heavy neutrino 
searches previously reported in \cite{nomadsg,nomadtau,nomadkarm}.
  \begin{figure}
 \begin{center}
   \mbox{\hspace{-.8cm}\epsfig{file=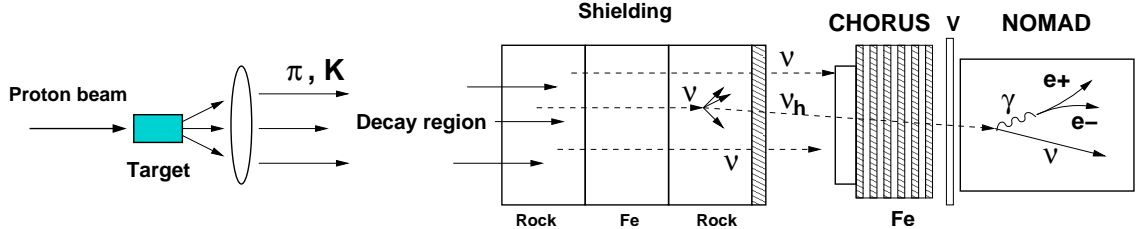,width=150mm}}
\end{center}
    \centering
\vspace{0.5cm}
    \caption{\em Schematic illustration of the experiment to search for the $\nuhd$ decay of a sterile neutrino.
If the $\nu_h$'s exist, they would be  produced  in $\nu_\mu$ interactions in the  CERN WANF $\nu_\mu$ neutrino beam line shielding.  The particles would penetrate the downstream shielding and be observed 
 in the NOMAD neutrino detector through their $\nuhd$  decays into a photon and a light neutrino followed by the 
photon conversion into  $e^+e^-$ pair in the NOMAD target.  The shaded region of the shielding  serves as a  dump of the thickness 
 about 9 $\lambda_{int}$ (interaction lengths) to absorb  neutrino shower components  
 which could escape absorption in the CHORUS detector matter, and generate background events in NOMAD.}
\label{setup}
\end{figure}

The CERN West Area Neutrino Facility (WANF) beam line \cite{wanf} 
provides an essentially pure $\nu_{\mu}$
beam for neutrino experiments. 
It consists of a beryllium
target irradiated by  450 GeV  protons from the CERN SPS.\ 
The secondary hadrons of a given sign 
are focused with two magnetic elements located in front of a 290 m long evacuated decay 
tunnel.\ Protons that do not interact in the target, secondary hadrons and 
muons that do not decay are absorbed by a 400 m thick shielding made of iron and 
earth.   The NOMAD detector is located at 835 m from the neutrino target.
The detector is described in detail in Ref.~\cite{nomad}.\ It consists of
a number of sub-detectors most of which are located inside a 0.4 T dipole 
magnet  with a  volume of 7.5$\times$3.5$\times$3.5 m$^{3}$:  
 an active target of 
drift chambers (DC) with a mass of 2.7 tons
(mainly carbon), and a total thickness
of about one radiation length
 followed by 
a transition radiation detector, a preshower detector, and an electromagnetic
calorimeter.  A hadron calorimeter and two
muon stations  are located just after the magnet coils.
A plane of scintillation counters,  $V$,  in front of the magnet was used to
veto  upstream neutrino interactions and muons incident on the detector. 

If the $\nu_h$ exists, it would be  produced in interactions
\begin{equation}
\nu_\mu + N \to \nu_h +X.
\label{reaction}   
\end{equation}
 of  muon neutrinos in  the WANF neutrino beam shielding.
If the $\nu_h$ is a relatively long-lived particle, the flux of $\nu_h$'s
would penetrate the downstream part of the shielding without significant attenuation  and be observed in  NOMAD through  $\nuhd$ decays followed by the decay photons conversion  into $\ee$ pairs 
in the NOMAD DC target, as schematically illustrated in Fig.\ref{setup}.
 The experimental signature  of $\nuhd$ decays would appear as 
an excess of isolated $\pair$ pairs above those expected from standard neutrino interactions. 
As the NOMAD searched for an excess of isolated  $\ee$ pairs 
 in their detector, the results obtained can be also used to constrain the radiative decay of the $\nu_h$.
Note, that the  search for an excess of isolated  photon events 
in neutrino interactions is complicated by the presence of a large background coming from the electromagnetic decays of neutral mesons (mostly $\pi^0, \eta, \eta' \to \gamma \gamma$). However, because  the experimental signature of these events is clean, they can be selected with significantly suppressed  background due to the excellent NOMAD capability for precise measurements of $\ee$ pairs, see for example \cite{nomadsg,nomadtau,nomadkarm}.   
\begin{figure}[htb!]
 \begin{center}
   \mbox{\epsfig{file=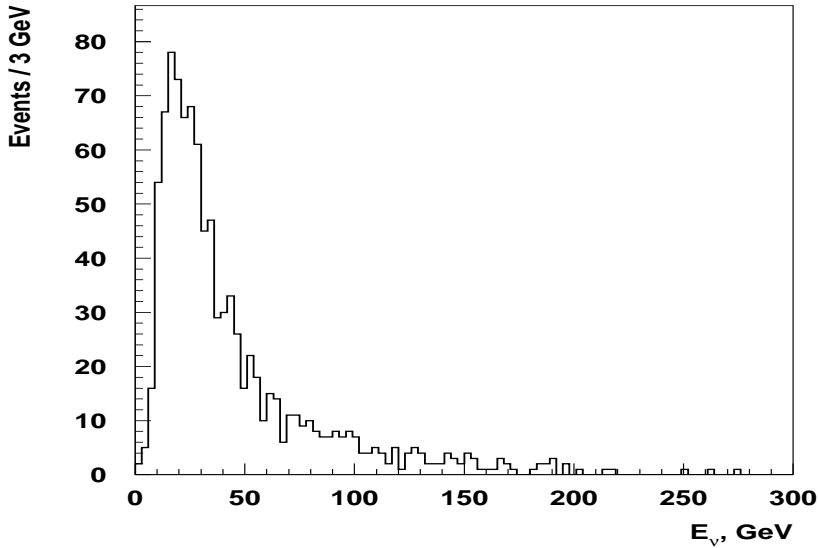, height=80mm,width=140mm}}
 \end{center}
\vspace{0.5cm}
    \centering
    \caption{\em Energy spectra of muon neutrinos that produce heavy neutrino with the mass of 50 MeV 
    in neutral current interactions in the WANF neutrino beam line shielding. }
 \label{numusp}
 \end{figure}
 
 To make quantitative estimates, 
we performed simplified simulations of the $\nu_h$ production \eqref{reaction} and decay in the NOMAD detector
schematically  shown in Fig. \ref{setup}.
The flux of $\nu_h$'s produced in the shielding was calculated by using the 
known flux of  $\nu_\mu$ beam in NOMAD \cite{flux}. In these simulation the $\nu_h$ production vertex was required 
to be located in the upstream part of the shielding excluding shaded region shown in Fig.\ref{setup}.
This end cap region  serves  as an additional dump  to absorb  neutrino hadronic shower components, which might escape absorption in the  matter of the CHORUS detector located downstream. 
   The choice of the dump thickness of 9 $\lambda_{int}$ (interaction lengths) allows to  fully 
 protect NOMAD from the leak of  secondary particles produced in the upstream neutrino interactions, 
 and suppress  background from punch-through $K_L^0$ and  $n$ interactions in NOMAD that could produce fake signal.
 More accurate  analysis would require  detailed simulations of the background 
  in the NOMAD detector, which is beyond the scope of this work. Note, however, that obtained 
limits are not very sensitive to the choice of the dump thickness within the range $\simeq 9 - 15$ $\lambda_{int}$.

\begin{figure}[htb!]
 \begin{center}
   \mbox{\epsfig{file=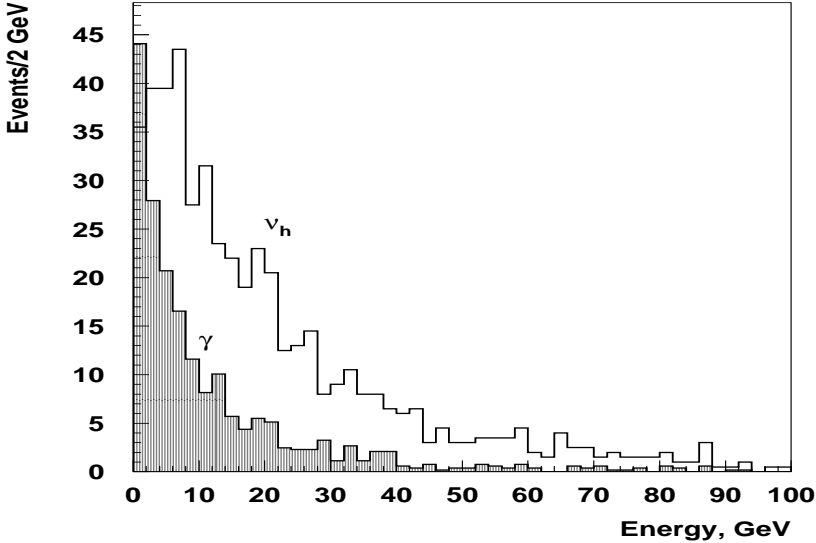, height=80mm,width=140mm}}
 \end{center}
\vspace{0.5cm}
    \centering
    \caption{\em Energy distributions  of $\nu_h$ neutrinos with the mass 50 MeV produced in the $\nu_\mu$NC interactions in the beam dump and   pointing towards the NOMAD fiducial area,  and  photons from the $\nuhd$ decay in the NOMAD target. 
    The spectra are normalized to a common maximum.}
 \label{energy}
 \end{figure}

The simulated energy spectrum of the incident $\nu_\mu$'s is  shown in Fig.\ref{numusp}. It is assumed 
that, similar to the case of neutrino mixing, the cross section   of heavy neutrinos production \eqref{reaction}  can be expressed in a model-independent way as follows:
\begin{equation}
 \sigma(\nu_\mu + N \to \nu_h + X) = \alpha_{\mu h} \sigma(\nu_\mu + N \to \nu_\mu + X)  f_{ph.s.}  
\label{crossec}
\end{equation}
 where $\sigma(\nu_\mu + N\rightarrow \nu_\mu + X)$ is the 
cross section for $\nu_\mu$NC interactions, $f_{ph.s.} = \sigma (m)/\sigma(0)$  is the  phase space 
 factor which takes into account  dependence on the $\nu_h$ mass, and   $\alpha_{\mu h}$ play a role of an effective coupling strength in new $\nu_h$-NC-like interactions.
The distribution 
of energies for $\nu_h$'s with momenta pointing to  the NOMAD fiducial area,  and  photons from the 
isotropic $\nuhd$ decay in the NOMAD target are shown  in Fig.\ref{energy}. 
Once the $\nus$ flux was known, the next step
was to calculate the $\pair$ spectrum based on  
the  $\nuhd$ decay rate.\ 
For a given flux  $\Phi(\nuh)$, the expected number of signal events from 
$\nuhd$ decays occurring within the fiducial length $L$ of the NOMAD 
detector located  at a distance $L'$ from the $\nu_h$ production vertex is given by 
\begin{equation}
N_{\nudecay} =\int  A \Phi(\nu_h) exp\bigl(-\frac{L'm_{\nu_h}}{p_{\nu_h}\tau_h}\bigr) 
\bigl[1-exp\bigl(-\frac{L m_{\nus}}{p_{\nuh}\tau_h}\bigr)\bigr] 
 \frac{\Gamma_{\gamma\nu}}{\Gamma_{tot}} P_\gamma \varepsilon_{\ee}  dE_{\nuh}dV
\label{rate}
\end{equation}
where $p_{\nuh}$ is the $\nu_h$ momentum, $\tau_h$ is its 
lifetime at the rest frame, $\Gamma_{\pair},~\Gamma_{tot}$
are the  partial and total $\nuh$-decay widths, respectively, $P_\gamma$ is the 
decay photon conversion probability, and  $\varepsilon_{\ee}$ is the $\pair$ pair reconstruction efficiency.
The acceptance $A$ of the NOMAD detector was calculated  tracing $\nus$'s
produced in the shielding  to the detector.
As an example for a mass $m_{\nus}= 50~\rm MeV$, $A=4.8\%$ and $\varepsilon_{\ee}\simeq 25\%$.\ 
It is assumed that the total rate $\Gamma_{tot}$ of the $\nu_h$ decays is dominanted by the 
 radiative decay $\nuhd$, see discussion in \cite{sng},
 hence the branching fraction  of the $\nuhd$  decay  is   
$BR(\nuhd) = \frac{\Gamma(\nuhd)}{\Gamma_{tot}}\simeq 1$.
 
 The NOMAD search for an excess of single photon events is  described in detail in Ref. \cite{nomadsg}.
Briefly, it  used  data collected with   of $5.1\times 10^{19}$ protons on target.
The strategy of the analysis was to identify $\ee$ candidates   by reconstructing in the DC isolated  $\pair$ pairs
with the total energy greater than 1.5 GeV  and invariant mass below 100 MeV, 
 that are accompanied  by no other activity in the detector. The measured rate 
of $\pair$ pairs was then compared to that expected from known sources.
 Two samples of  signal photon candidate events were selected in NOMAD  among the total 
 number of 1.44$\times 10^6$ $\nu_\mu$CC reaction recorded \cite{nomadsg}. The first sample corresponds to photons 
emitted at any angle $\Theta_{\nu \gamma}$ with respect to the primary neutrino beam direction, while the second one, corresponding to photons produced  in the forward direction, was selected by using  the cut $\zeta < 0.05$ GeV on   variable $\zeta = E_\gamma (1-cos\Theta_{\nu \gamma})$. 
After applying all selection criteria,  155 and 78  candidate events, with a predicted background of 
$129.2 \pm 8.5 \pm 3.3$ and 76.6$\pm$ 4.9 $\pm$ 1.9, were observed in two regions, respectively.
These results are found to be consistent with the background expectations and  yield an excess of 25.8 $\pm$15.5 
and 1.4 $\pm$10.3 events, respectively. The measured spectra are also found to be in good agreement with 
prediction. Hence, no evidence for an excess of single photons produced in $\nu_\mu$ neutrino 
interactions   has been observed and the corresponding upper limits of $<$ 51 events and $<$ 18 events at 90\% CL for the number of single photon excess events in each sample were obtained, respectively. 
\begin{figure}
 \begin{center}
   \mbox{\epsfig{file=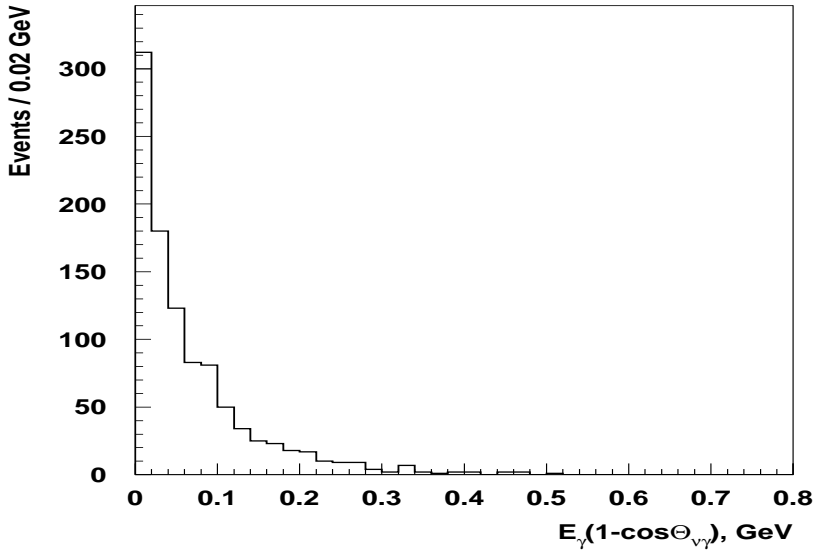, height=80mm,width=140mm}}
    \end{center}
\vspace{0.5cm}
    \centering
    \caption{\em Distribution  of variable $E_\gamma (1-cos\Theta_{\nu \gamma})$ for photons from the decay $\nuhd$ of the $\nuh$ with the mass 50 MeV produced in the $\nu_\mu$NC interactions in the SPS neutrino beam dump and  
decaying in  the NOMAD fiducial volume.}
  \label{gan}
 \end{figure}
 
Using  Eq.(\ref{rate}) and these upper limits, we can then determine the $90\%$ CL
upper limit for the corresponding coupling $\alpha_{\mu h}$. 
The distribution of variable $\zeta$ for decay photons in NOMAD is shown in Fig.\ref{gan}. As the $\nu_h$'s 
arrive at NOMAD from the source located at a relatively far average distance of $\simeq 70$ m, most of the decay photons 
are produced in the forward direction, at a small values of $\zeta$ and angle $\Theta_{\nu \gamma}$. In this case,   
 the best bounds are obtained by selecting signal events with the  cut $\zeta < 0.05$ GeV. 
The corresponding 90$\%$ CL exclusion region in the
$\alpha_{\mu h}$ vs $\tau_h$ plane shown in Fig. \ref{plot} together with the  result obtained with the CHORUS target (see below) is  calculated by using the relation $N_{\nuhd} < 18 $ events. Our result is sensitive to a coupling $\alpha_{\mu h} \gtrsim 10^{-3}$, and corresponds to the $\nu_h$ lifetime region  $\tau_h \simeq 10^{-10}-10^{-9}$ s.
Over most of this region, the $\nu_h$  lifetime is sufficiently long, that $Lm_{h}/p_{h}\tau_h \ll L'm_{h}/p_{h}\tau_h \ll 1$. 

For the $\nu_h$ lifetimes shorter than $\tau_h \simeq 10^{-9}$ s, new constraints on $\nu_h$ properties can be  obtained by considering the $\nu_h$ production in the  muon spectrometer matter (mainly iron)  
of the CHORUS detector located just about 10 m upstream  of the NOMAD at the same neutrino 
beam line, as shown in Fig.\ref{setup}.
The CHORUS detector, specifically designed to search for $\nu_\mu - \nu_\tau$ oscillations at the SPS
$\nu_\mu$ neutrino beam,  is a hybrid setup which combines a nuclear emulsion target 
with various electronic detectors such as trigger hodoscopes, a scintillating fibre 
tracker system, a hadron spectrometer, electromagnetic and hadronic calorimeters, 
and the muon spectrometer \cite{chorus}. 
The muon spectrometer consists of seven tracking sections interleaved by six magnets.
Each magnet is made from twenty iron disks, each of 3.75 m in diameter and 2.5 cm thick,  with scintillator planes interspersed. The total weight of this Fe-target is about 260 t.
\begin{figure}
 \begin{center}
   \mbox{\epsfig{file=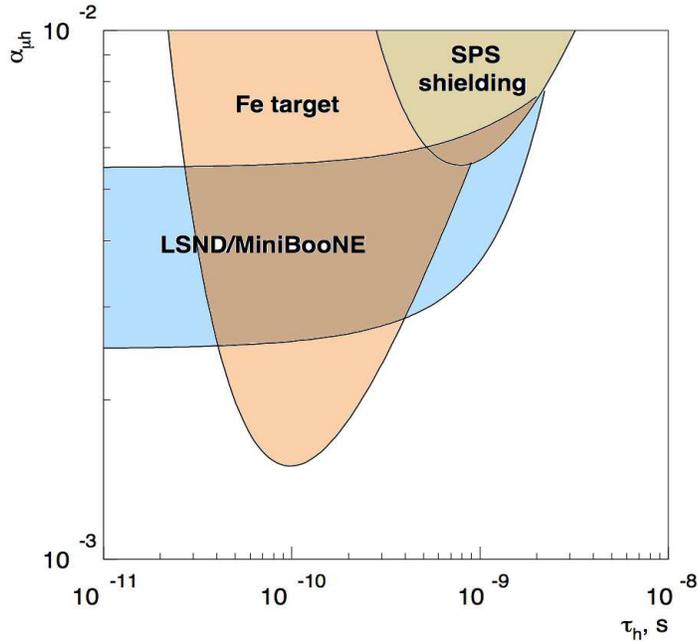,height=90mm,
width=120mm}}
\end{center}
    \centering
    \caption{\em  The  90\% $CL$ exclusion regions in the ($\tau_h, \alpha_{\mu h}$)- plane obtained from the NOMAD results 
\cite{nomadsg}    for the case of the $\nu_h$ production in the SPS neutrino beam shielding   and in the iron of the muon spectrometer 
    of the CHORUS detector (Fe target). The area of the allowed LSND/MinibooNE parameter space corresponding to the $\nu_h$ mass of 50 MeV is also shown.}
  \label{plot}
  \end{figure}
  
Similar to above considerations, the expected number of heavy neutrino $\nuhd$ decays in the fiducial volume of the NOMAD detector can be estimated by using expression \eqref{rate}, and  taking into account  the corresponding number of heavy neutrinos produced in the primary neutral current interaction in the CHORUS iron target. 
In this estimate we required  the $\nu_h$ production to occur in the first three upstream  sections of the CHORUS muon
spectrometer, assuming that the hadronic showers from the neutrino interaction are completely absorbed in the remaining three  downstream sections served as a "beam dump" with the total thickness of 9 $\lambda_{int}$. 
This requirement is necessary to suppress background from $K_L^0$ and $n$ interactions discussed above,
 and also to avoid rejection of signal event by the NOMAD VETO system, when a secondary energetic hadron trigger the NOMAD veto counter V. The simulations show that due to the shorter distance 
to the Fe target the angular distribution of the decay photons converted in the  NOMAD DC target is wider 
in comparison with the case, considered above, and the most restrictive limits are derived  by using 
the NOMAD 90\% CL upper limit of $<$ 51 events obtained for the case of the wider $\zeta$
distribution of the single photon events. 

The final 90\% CL upper limit curves  in the 
($\tau_h; \alpha_{\mu h}$) plane 
are shown in Fig. \ref{plot} together with the LSND-MiniBooNE $\nu_h$ parameter space calculated for the $\nu_h$ mass of 50 MeV.
The attenuation of the $\nu_h$ flux due to absorption in the shielding is found to be negligible. For example,  
our results can also be used to constrain the 
magnitude of the transition magnetic moment ($\mu_h$) of the hypothetical $\nu_h$, see for more details Ref.\cite{sng}. 
The estimated attenuation of the $\nu_h$ - flux due to $\nu_h$ electromagnetic  interactions 
with matter in this case is found to be negligible, since for the $\nu_h$ lifetime in the range $10^{-11}\lesssim \tau_h \lesssim 10^{-9}$ s, 
the $\mu_h$ ranges from  $\simeq 10^{-8} \mu_B$ to $\simeq 10^{-9} \mu_B$ (here $\mu_B$ is the Bohr magneton), and 
 the heavy neutrino  mean free path in iron is more than  10 km, as compared with the iron 
and earth shielding total length of 0.4 km used in the SPS neutrino  beam.

In summary, in this work we study possible manifestations of the radiative decays $\nuhd$ of a heavy sterile neutrino
$\nuh$ in the CERN WANF neutrino beam. The existence of the $\nuh$ was proposed for the explanation of the anomalous excess events observed in the LSND and MiniBooNE experiments. It is  assumed that $\nu_h$'s are dominantly 
produced  in new $\nu_\mu$NC-like reactions in order to evade recent constraints  from the $\nu_h$ searches in 
reactions induced by $\nu_\mu$CC interactions. We study the $\nu_h$ production 
 occurring  either in  the CERN WANF neutrino beam line shielding, or 
 in the iron of the muon spectrometer of the CHORUS detector located upstream  of the NOMAD and followed
 by the radiative decay $\nuhd$ with the subsequent conversion of decay photons into $\ee$ pairs
 in the NOMAD active target. Using sensitive limits from the  NOMAD experiment on single photon production in neutrino interactions, we  derive new constraints on the $\nu_h$ properties.  The obtained results allow to exclude  $\nu_h$'s  of \eqref{param} for the 
lifetime values $3\times 10^{-11}\lesssim \tau_h \lesssim 5 \times 10^{-10}$ s, while still 
leaving open two bands for  the existence
of $\nu_h$ with the lifetime either $\tau_h \simeq 10^{-9}$ s or  $ \tau_h \lesssim 3 \times 10^{-11}$ s. 
We have  demonstrated a significant  potential  of the discussed experimental approach for the future more sensitive searches for the $\nu_h$ at the high intensity neutrino facilities.
 For example, more restrictive constraints are expected to be obtained  from the search for the $\nuhd$ decays from the 
 $\nu_h$ produced in  $\nu_\mu$ interactions in the future LAr experiment \cite{proposal},
 aiming to  check the origin of the LSND/MiniBooNE excess events at a  neutrino facility at CERN \cite{2lar1,2lar2}.

I would like to thank D. Autiero,  D.S. Gorbunov, A. Guglielmi, L. Camilleri, M.M. Kirsanov, N.V. Krasnikov, L. Di Lella, 
V.A. Matveev, F. Pietropaolo, A. Rubbia, C. Rubbia, and S. Mishra    for their interest, and valuable discussions and/or comments.  The help of D.S. Gorbunov and   M.M. Kirsanov is greatly appreciated.

\end{document}